\documentclass[review]{elsarticle}

\usepackage{lineno,hyperref,amsmath,bm}
\usepackage{multirow,adjustbox}
\biboptions{super,sort&compress}
\journal{}

\begin{document}

\begin{frontmatter}

\title{Valence and core excitons in solids from velocity-gauge real-time TDDFT with range-separated hybrid functionals: An LCAO approach}
%\tnotetext[mytitlenote]{Fully documented templates are available in the elsarticle package on %\href{http://www.ctan.org/tex-archive/macros/latex/contrib/elsarticle}{CTAN}.}

\author{C. D. Pemmaraju\corref{ca}}
\cortext[ca]{Corresponding author}
\ead{dasc@slac.stanford.edu}

\address{Stanford Institute for Materials and Energy Sciences, SLAC National Accelerator Laboratory, Menlo Park, CA 94025, USA}





\begin{abstract}
An atomic-orbital basis set framework is presented for carrying out velocity-gauge real-time time-dependent density functional theory (TDDFT) simulations in periodic systems employing range-separated hybrid functionals.  Linear optical response obtained from real-time propagation of the time-dependent Kohn-Sham equations including nonlocal exchange is considered in prototypical solid-state materials such as bulk Si, LiF and monolayer hexagonal-BN. Additionally core excitations in monolayer hexagonal-BN at the B and N K-edges are investigated and the  role of long-range and short-range nonlocal exchange in capturing valence and core excitonic effects is discussed. Results obtained using this time-domain atomic orbital basis set framework are shown to be consistent with equivalent frequency-domain planewave results in the literature. The developments discussed lead to a time-domain generalized Kohn-Sham TDDFT implementation for the treatment of core and valence electron dynamics and light-matter interaction in periodic solid-state systems. 
\end{abstract}


\end{frontmatter}

%\linenumbers

\section{Introduction}
Time-Dependent Density Functional Theory (TDDFT)~\cite{Runge1984,Marques2012a,Ullrich2012} which represents a generalization of density functional theory (DFT)~\cite{Hohenberg1964,Kohn1965} to time-dependent systems is  well established as a computationally efficient first-principles methodology for accessing excited state properties in both molecular and solid state materials.  While linear-response TDDFT~\cite{CASIDA1995} (LR-TDDFT) has been a standard feature in first-principles quantum chemistry codes for over two decades~\cite{Casida2012,Burke2005}, real-time TDDFT (RT-TDDFT)~\cite{Yabana1996,Bertsch2000} has grown significantly in prominence within the last ten years driven primarily by the need for theoretical developments to complement experimental efforts employing time-domain laser spectroscopies. To date, a number of implementations of RT-TDDFT have been reported  ~\cite{Yabana1996,Bertsch2000,Tsolakidis2002,Takimoto2007,Meng2008,Lopata2011,Andrade2012,Wang2013,Krieger2015,Goings2016,Provorse2016,Yost2017,salmon,Pemmaraju2018,Jia2018,Lian2018} with applications to both molecular and solid-state systems. In particular, for the  non-perturbative treatment of light-matter interactions in periodic solids especially outside the linear-response regime, the velocity gauge (VG) formulation of RT-TDDFT originally proposed by Yabana and Bertsch~\cite{Yabana1996,Bertsch2000} has proven useful and numerical implementations of this scheme employing real-space-grids\cite{Yabana2012,salmon,Andrade2012}, full-potential linearized augmented planewaves (FP-LAPW) \cite{Dewhurst2004,Krieger2015} and atomic-orbital basis sets~\cite{Pemmaraju2018,Lian2018} have been demonstrated in connection with a wide range of applications.

Analogously to the case of DFT for ground state properties, the predictive accuracy of TDDFT for excited states depends on the functional approximation utilized to describe exchange-correlation (XC) effects characterizing the many-electron system~\cite{Ullrich2012,Marques2012a}.  A large body of literature investigating exact properties of the time-dependent XC potential in TDDFT as well as a hierarchy of practical XC kernel approximations for routine TDDFT simulations has been developed over the years~\cite{Burke2005,Marques2012a,Ullrich2012,Casida2012,Baer2010,Baer2018}.  The Adiabatic Local Density Approximation (ALDA) wherein the XC potential is both a space and time-local multiplicative quantity represents the simplest XC approximation within  Kohn-Sham TDDFT~\cite{Marques2012a,Ullrich2012}.  The ALDA was widely utilized in the early days of TDDFT but is characterized by a number of short-comings. Among the most well-known failures of the ALDA are its inability to describe charge transfer excitations in molecules and the lack of exciton binding in solids~\cite{Marques2012a,Ullrich2012, Baer2010, Casida2012,Kronik2016,Turkowski2017,Kummel2017}. Since the advent of Generalized Kohn-Sham (GKS)~\cite{Seidl1996,Baer2010,Perdew2017} theory over two decades ago, the ALDA has been largely superseded within molecular quantum chemistry by a wide array of hybrid-DFT functionals and the corresponding non-local XC kernels that provide much improved accuracy in both linear-response and real-time TDDFT applications~\cite{Marques2012a,Ullrich2012, Baer2010, Casida2012,Kronik2016,Turkowski2017,Kummel2017}. These developments have recently been placed on a formally rigorous footing by Baer and Kronik within \textit{Generalized Kohn-Sham} TDDFT (GKS-TDDFT)~\cite{Baer2018}. 

In solid-state systems, the importance of long-range nonlocal exchange for capturing excitonic effects within TDDFT has been widely discussed~\cite{Botti2004,Marques2012a,Ullrich2012, Yang2015, Refaely-Abramson2015a, Kronik2016, Turkowski2017}, but practical implementations of nonlocal kernels grounded in GKS theory have taken longer to emerge primarily due to the computational cost. In recent years, the development of efficient algorithms for nonlocal exchange within planewave basis set implementations has led to wide spread utilization of GKS XC approximations within ground state DFT simulations of periodic solids~\cite{Marsman2008,Betzinger2010,Bylaska2011,Lin2016,Barnes2017}.  Concomitantly, a number of atom-centered Gaussian-type orbital (GTO)~\cite{Causa1988,Heyd2003,Tymczak2005a,Hutter2014} or Numerical atomic orbital (NAO)~\cite{Fernandez2003,Shang2011,Levchenko2015,Qin2015} based condensed matter DFT platforms have also been developed to take advantage of GKS XC functionals in periodic systems. Several notable studies have extended these developments to TDDFT calculations of excited states in solids at the GKS level of theory~\cite{Paier2008,Hutter2014,Yang2015,Sato2015,Refaely-Abramson2015a,Turkowski2017,Jia2018a}. In particular recent studies employing tuned range-separated hybrid functionals within planewave basis set implementations demonstrated good predictive accuracy for linear-response TDDFT with regards to excitonic effects in condensed matter systems~\cite{Refaely-Abramson2015a}. 

Taking advantage of the aforementioned developments,  in this article a linear combination of atomic orbitals (LCAO) approach is described for performing velocity gauge real-time TDDFT (VG-RT-TDDFT)~\cite{Yabana1996,Yabana2012,Pemmaraju2018} simulations of light-matter interaction and electron dynamics in periodic solids employing range-separated hybrid DFT functionals.  LCAO basis sets offer distinct computational efficiencies with regards to the treatment of localized core-electrons and the linear scaling with system size of the nonlocal Fock matrix relevant to GKS functionals~\cite{Tymczak2005a,Levchenko2015,Qin2015}. The present approach aims to utilize these efficiencies to enable GKS VG-RT-TDDFT simulations for light-matter interactions spanning near infrared (NIR) to soft X-ray energies while treating valence and core electronic excitations on the same footing. Such a development is useful in two respects: Firstly, whereas the vast majority of RT-TDDFT simulations in solids to date, with some notable exceptions~\cite{Sato2015,Yost2017,Jia2018a}, have been limited to ALDA or equivalent approximations, a framework such as the one discussed here could enable routine  VG-RT-TDDFT simulations at the GKS level of theory. Secondly, the unified treatment of core and valence excitations represents a useful first-principles theoretical complement to emerging ultrafast spectroscopic studies in solids utilizing high-harmonic generation (HHG) and X-ray free electron laser (XFEL) light sources to probe  electron dynamics from the point of view of core excitations~\cite{Schultze2014,Zhang2015,Moulet2017}.     
\section{Numerical implementation details}
The developments reported in this article extend an LCAO VG-RT-TDDFT framework at the ALDA level of theory that has been previously described in detail~\cite{Pemmaraju2018}. Therefore aspects related to generalizing the methodology beyond ALDA towards employing nonlocal XC functionals form the main focus of the following discussion.  The primary equations of interest are the time-dependent Generalized Kohn-Sham (TDGKS) equations for electron dynamics in the velocity gauge (VG)~\cite{Baer2018,Sato2015}:
\begin{align}\label{vgtdks}
&\imath\hbar\frac{\partial}{\partial t}\tilde{\psi}_i(\overrightarrow{r},t)=\hat{\tilde{H}}_{GKS}\tilde{\psi}_i(\overrightarrow{r},t)
\end{align} 
wherein $\tilde{\psi}_i(\overrightarrow{r},t)$ represent the VG time-dependent GKS orbitals and the VG Hamiltonian $\hat{\tilde{H}}_{GKS}$ given by
\begin{align}\label{HGKS}
\hat{\tilde{H}}_{GKS}=\frac{1}{2m}\left[  \overrightarrow{p} +\frac{e}{c}\overrightarrow{A}(t)\right]^2  + \hat{\tilde{V}}_{ion}+\int d\overrightarrow{r}^\prime \frac{e^2}{|\overrightarrow{r}-\overrightarrow{r}^\prime|}n(\overrightarrow{r}^\prime,t) + \hat{V}_{XC}[\rho(\overrightarrow{r},\overrightarrow{r}',t)]
\end{align}
includes the kinetic term incorporating coupling to time-dependent external fields via the vector potential $\overrightarrow{A}(t)$, the VG electron-nuclear interaction $\hat{\tilde{V}}_{ion}$, the Hartree potential and in general a non-multiplicative XC operator $\hat{V}_{XC}$ which is a functional of the instantaneous single-particle density matrix $\rho(\overrightarrow{r},\overrightarrow{r}',t)$~\cite{Baer2018}.  Propagating the TDGKS equation~\ref{vgtdks} in time yields the time-dependent density matrix,
\begin{equation*}
\rho(\overrightarrow{r},\overrightarrow{r}',t)=\sum_{i}^{N_{occ}} \tilde{\psi}_i(\overrightarrow{r},t)\tilde{\psi}_i^*(\overrightarrow{r}',t)
\end{equation*}
density,
\begin{equation*}
n(\overrightarrow{r},t)=\rho(\overrightarrow{r},\overrightarrow{r},t)=\sum_{i}^{N_{occ}}|\tilde{\psi}_i(\overrightarrow{r},t)|^2
\end{equation*}
and macroscopic current
\begin{equation}\label{macI}
\overrightarrow{I}(t)=-\frac{e}{\Omega}\int_{\Omega}d\overrightarrow{r}\overrightarrow{j}(\overrightarrow{r},t).
\end{equation}
The time-dependent current density $\overrightarrow{j}(\overrightarrow{r},t)$ in (\ref{macI}) is defined as 
\begin{equation}
\overrightarrow{j}(\overrightarrow{r},t)=\sum_{i}\frac{e}{2m} \left\lbrace \tilde{\psi}^*_i(\overrightarrow{r},t) \overrightarrow{\pi}\tilde{\psi}_i(\overrightarrow{r},t) + c.c \right\rbrace
\end{equation}
and includes the generalized momentum
\begin{equation}
\overrightarrow{\pi} = \frac{m}{\imath\hbar} [\overrightarrow{r}, \hat{\tilde{H}}_{GKS}].
\end{equation}
From the above, observables related to the time-dependent density or current are therefore readily available and frequency domain quantities can be calculated through Fourier transforms\cite{Marques2012a,Bertsch2000,Yabana2012,Pemmaraju2018}.

In the present implementation, the XC potential $\hat{V}_{XC}$ in $\hat{\tilde{H}}_{GKS}$ is obtained from employing the range-separated hybrid (RSH) functional~\cite{Baer2010,Refaely-Abramson2015a} form for the exchange-correlation energy $E_{XC}$:
\begin{align}\label{ERSH}
E_{XC} = \alpha E^\mathrm{SR}_{HFX} + (1-\alpha) E^\mathrm{SR}_{LDAX} +  (\alpha+\beta)E^\mathrm{LR}_{HFX} + [1-(\alpha+\beta)]E^\mathrm{LR}_{LDAX}
+E_{LDAC}
\end{align}
where $E^\mathrm{SR}_{HFX}, E^\mathrm{LR}_{HFX}$ represent short-range (SR) and long-range (LR) nonlocal Hartree-Fock exchange (HFX) respectively, $E^\mathrm{SR}_{LDAX},E^\mathrm{LR}_{LDAX}$ represent  short-range and long-range LDA exchange (LDAX)~\cite{Toulouse2004} respectively and $E_{LDAC}$ represents the LDA correlation energy~\cite{Perdew1981}. The coefficients $\alpha, \beta$ determine the fractions of each of the above SR and LR quantities within the total XC energy.  The SR and LR forms of the HF and LDA exchange energies are obtained by partitioning the Coulomb operator as:
\begin{align}\label{RSHC}
\frac{1}{|\overrightarrow{r}-\overrightarrow{r}^\prime|} = \frac{ \alpha + \beta~\mathrm{erf}(\omega |\overrightarrow{r}-\overrightarrow{r}^\prime| )}{|\overrightarrow{r}-\overrightarrow{r}^\prime|} + \frac{1-\lbrace \alpha + \beta~\mathrm{erf}(\omega |\overrightarrow{r}-\overrightarrow{r}^\prime| )\rbrace}{|\overrightarrow{r}-\overrightarrow{r}^\prime|}
\end{align}
The parameters $\alpha, \beta$ and the range-separation parameter $\omega$ together determine the overall mix of LDAX and HFX at different length-scales~\cite{Baer2010,Refaely-Abramson2015a}.  In this work, for $E^\mathrm{SR}_{LDAX}$, the Coulomb attenuated form of the LDA exchange due to Toulouse \textit{et al}~\cite{Toulouse2004} adapted from the implementation within the libxc library~\cite{Lehtola2018} is employed. For the evaluation of HFX, a standard real space approach for periodic systems involving the calculation of four-center electron repulsion integrals (ERIs) between atomic orbital basis functions distributed over an extended auxiliary supercell is used~\cite{Causa1988,Levchenko2015,Qin2015}. Accordingly, matrix elements  of the HFX  operator $\hat{X}^{\sigma}$ are first constructed in real-space as:
\begin{align}\label{HFXM}
X^{\sigma}_{ij}(\mathbf{R}) = \sum_{\mathbf{R}_1\mathbf{R}_2}\sum_{p,q} D^{\sigma}_{pq}(\mathbf{R}_2-\mathbf{R}_1)[\phi_i^\mathbf{0} \phi_p^{\mathbf{R}_1}||\phi_q^{\mathbf{R}_2}\phi_j^{\mathbf{R}}]
\end{align}
 where $\sigma$ is the spin index, $\mathbf{0}$ represents the reference unitcell and $\mathbf{R},\mathbf{R}_1,\mathbf{R}_2$ are lattice vectors spanning the auxiliary supercell about the reference cell, $\phi_n^{\mathbf{L}}$ represents a basis function with index $n$ within the lattice cell at $\mathbf{L}$, $D^{\sigma}_{nm}(\mathbf{L})$ is the real-space density matrix (DM) element connecting orbitals $\phi_n^{\mathbf{0}},\phi_m^{\mathbf{L}}$ and the quantity within square brackets is an ERI expressed in the general case of a range-separated Coulomb operator as
 \begin{align}\label{ERI}
 [\phi_i^\mathbf{0} \phi_p^{\mathbf{R}_1}||\phi_q^{\mathbf{R}_2}\phi_j^{\mathbf{R}}] = \int \int  d\overrightarrow{r} d\overrightarrow{r}^\prime \frac{\phi_i^\mathbf{0*}(\overrightarrow{r})\phi_p^{\mathbf{R}_1*}(\overrightarrow{r})\mathrm{erf}(\omega |\overrightarrow{r}-\overrightarrow{r}^\prime| )\phi_q^{\mathbf{R}_2}(\overrightarrow{r}^\prime)\phi_j^{\mathbf{R}}(\overrightarrow{r}^\prime)}{|\overrightarrow{r}-\overrightarrow{r}^\prime|}
 \end{align}
The momentum space HFX matrix is subsequently obtained at points within the first Brillouin zone (BZ) through Fourier transformation as
 \begin{align}\label{HFXK}
 \mathcal{X}^{\sigma}_{ij}(\mathbf{k}) = \sum_\mathbf{R} e^{\imath \mathbf{k} \mathbf{R}} X^{\sigma}_{ij}(\mathbf{R})
 \end{align}
In the case of range-separated hybrid functionals exchange contributions from both the standard $1/r$ and range-separated $\mathrm{erf}(\omega r)/r$  Coulomb operators are calculated separately and weighted according to the $\alpha$ and $\beta$ parameters in equation \ref{RSHC} to form the total XC contribution within the GKS Hamiltonian~\cite{Baer2010,Refaely-Abramson2015a}. The HFX like contribution to the total XC energy  is easily evaluated in the real-space approach as 
\begin{align}
E_{HFX}=-\frac{1}{2} \sum_{\sigma} \sum_{ij \mathbf{R}} D^{\sigma}_{ij}(\mathbf{R}) X^{\sigma}_{ij}(\mathbf{R})
\end{align}

Within the above scheme, the primary computationally demanding tasks are the evaluation of the ERIs in equation~\ref{ERI} and the summations in equation~\ref{HFXM}. The ease of computing the ERIs strongly depends on the choice of orbital basis functions employed~\cite{Causa1988,Qin2015,Levchenko2015}. Gaussian type orbitals (GTOs) which are standard in molecular quantum chemistry feature analytical expressions for the ERIs that facilitate rapid computation~\cite{Helgaker1995,}. ERI schemes for other choices such as Slater type orbitals~\cite{TeVelde2001} and fully numerical atomic orbitals (NAO)~\cite{Levchenko2015} have also been reported in the literature. The present VG-RT-TDDFT implementation is based on a development version of the SIESTA~\cite{Soler2002} code which features NAO basis functions in its standard operating mode. While the direct evaluation of ERIs between NAO basis functions is possible as has been demonstrated by Levchenko et al~\cite{Levchenko2015}, the numerical integration schemes necessary for NAOs are typically slower compared to those for GTO basis functions and furthermore require significant algorithmic infrastructure to be built from the ground up.  On the other hand, ERI techniques for GTOs have been under development for over five decades and are readily accessible through modern open source libraries such as libint~\cite{Libint2} and libcint~\cite{Sun2015}. Therefore, for the evaluation of ERIs a hybrid approach based on expanding individual NAOs over a small number of GTOs is adopted in this instance. The method is closely related to the one already reported by Qin et al~\cite{Qin2015} and consists of the following steps:\\ 
(i) NAO basis functions for the system of interest are  generated using standard procedures within SIESTA~\cite{Soler2002} at the DFT LDA level.\\  
(ii) Each NAO basis function is fit to a small number (typically 3-4) of primitive GTOs using a combination of Levenberg-Marquardt~\cite{Press2007} and simplex~\cite{Press2007} optimization. Thus each NAO is approximated as a contracted GTO. \\
(iii) The GTO exponents and contraction coefficients determined in (ii) are used in conjunction with the integral evaluation scheme for real-spherical GTOs provided by the libcint~\cite{Sun2015} library to calculate effective ERIs between NAOs.\\
As shown by Qin et al~\cite{Qin2015} and also later on in this article, the above scheme leads to hybrid DFT simulations in good agreement with planewave basis set approaches where the HFX matrix is evaluated through momentum space algorithms~\cite{Marsman2008}. Once an ERI calculation procedure is selected the summations over direct lattice sites implied in equation~\ref{HFXM} need to be carried out. For a detailed discussion of the real-space convergence properties of the HFX matrix and related integral screening algorithms the reader is referred to the extensive existing literature on the subject~\cite{Causa1988, Heyd2003,Levchenko2015}. Briefly, for a short-ranged $\mathrm{erfc}(\omega r)/r$ form of the Coulomb interaction, real-space convergence of the HFX matrix $X^{\sigma}_{ij}(\mathbf{R})$ is straightforward and practically achieved by extending the lattice sums in (\ref{HFXM}) over an auxiliary supercell whose dimensions are consistent with the length-scale of the interaction set by the range-separation parameter $\omega$. For an unscreened $1/r$ form of the Coulomb interaction the HFX matrix is in general long-ranged and the real-space decay of the density matrix primarily determines the range of the auxiliary lattice summations necessary for convergence~\cite{Causa1988,Heyd2003,Levchenko2015}. In practice therefore, when long-range HFX is included, convergence has to be tested on a case by case basis. In the current implementation, the size of the real-space auxiliary supercell is included as a tunable parameter to allow for convergence to be verified to within a chosen tolerance level or to the extent permitted by computational feasibility. Because of the correspondence between real-space and momentum-space, similar considerations apply in reciprocal-space algorithms where the size of the $\mathbf{q}$-point grid used for evaluating HFX is analogous to the size of the auxiliary supercell in real-space methods~\cite{Marsman2008,Paier2008,Levchenko2015}.  
Once the HFX matrix is constructed as above it is combined with the LDAX and correlation contributions according to the recipe in equation~\ref{ERSH}  to form the full range-separated hybrid XC matrix that directly enters the VG-GKS Hamiltonian in equation~\ref{HGKS}. Because of the unitary invariance of the HFX matrix no additional complications are introduced by transformation to the velocity gauge. The VG-TD-GKS equations are then evolved using a standard Crank-Nicholson~\cite{Crank1947} scheme with a predictor-corrector step~\cite{Takimoto2007,Pemmaraju2018}. 
\begin{table}[htbp]
\begin{center}
	\begin{tabular}{ |c|c|c| }
		\hline
		\multicolumn{3}{|c|}{Computational parameters for bulk solids common to LCAO and PAW} \\
		\hline
		& bulk Si &  bulk LiF \\
		\hline
		lattice constant   & 5.429 \AA   & 4.017 \AA \\
		\hline
		LDA pseudopotential & Si: [Ne]3$s^2$,3$p^2$ & Li: [He]2$s^1$\\
		valence configuration &  & F:[He]2$s^2$,2$p^5$\\
		\hline
		DFT SCF $\mathbf{k}$-point grid & $\Gamma-6\times6\times6$ & $\Gamma-8\times8\times8$\\
		\hline
		IPA optics $\mathbf{k}$-point grid &$\Gamma-30\times30\times30$&$\Gamma-24\times24\times24$\\
		\hline
		\multicolumn{3}{|c|}{LCAO specific parameters} \\
		\hline
		\multirow{2}{*}{basis set ($nl$-$\zeta$)} &  Si: $3s$-2,$4s$-1,$3p$-2,$3d$-2  & Li: $2s$-2, $2p$-2  \\
		&    & F: $2s$-2, $2p$-2, $3d$-1  \\
		\hline
		Real-space mesh-cutoff & 364 Ry & 526 Ry  \\
		\hline
		HFX auxiliary supercell & $6\times6\times6$ & $8\times8\times8$\\
		\hline
		VG-RT-TDDFT $\mathbf{k}$-point grid & $\Gamma-30\times30\times30$ & $\Gamma-24\times24\times24$\\
		\hline
		VG-RT-TDDFT time step & 0.08 a.u & 0.08 a.u\\
		\hline
		\multicolumn{3}{|c|}{PAW specific parameters} \\
		\hline
		planewave cutoff  & 400 eV & 400 eV \\
		\hline
		HFX $\mathbf{q}$-point grid & $6\times6\times6$ & $8\times8\times8$\\
		\hline
	\end{tabular}
\end{center}
\caption{Computational parameters used for bulk Si and LiF simulations. To model the interaction with ionic cores, norm-conserving LDA pseudopotentials generated using the Troullier-Martins~\cite{Troullier1991} scheme are employed within LCAO runs while PAW simulations use standard LDA-PAW datasets provided with the VASP~\cite{Hafner2008} distribution. The same reference valence electronic configurations are used in both cases.  $\Gamma$-centered $\mathbf{k}$-point grids of the same dimensions are used in both codes to converge the DFT self-consistent field (SCF) and finer $\mathbf{k}$-point grids are employed to subsequently calculate linear optical response at the level of the independent particle approximation (IPA). The same $\mathbf{k}$-point grid dimensions used for IPA optical response are also used within LCAO based VG-RT-TDDFT simulations. Auxiliary supercell dimensions used to calculate HFX in the real-space LCAO approach are chosen to be consistent with $\mathbf{q}$-point grids used to calculate HFX in momentum space within the PAW method. The LCAO basis set is indicated using $nl$-$\zeta$ notation where $n,l$ are principal and azimuthal quantum numbers respectively and $\zeta$ is the number of functions of each $nl$ type. }\label{tab1}
\end{table}
\section{Results}
In the following, results obtained using the LCAO GKS implementation described above are presented. A number of prototypical materials such as the intermediate gap covalent semiconductor bulk Si, the wide band gap ionic insulator bulk LiF and the low-dimensional insulator monolayer hexagonal-BN (h-BN) are considered. Each one of the above systems requires a significantly different fraction of long-range asymptotic HFX and the results therefore span a broad parameter range in the context of RSH DFT and TDDFT~\cite{Baer2010,Refaely-Abramson2015a}. The results are presented in two distinct steps: Firstly, the validity of the LCAO GKS implementation is verified at the level of GKS-DFT band structure calculations by comparison with corresponding planewave projector augmented wave (PAW)~\cite{Blochl1994,Kresse1996,Marsman2008} simulations also carried out as a part of this work. Subsequently, GKS VG-RT-TDDFT results obtained with the LCAO scheme are compared against pre-existing planewave results in the literature and/or experiment.  The LCAO simulations are performed with an in-house development version~\cite{Pemmaraju2018} of the SIESTA~\cite{Soler2002} code. The PAW calculations are carried out using the VASP~\cite{Kresse1996,Hafner2008} package (version 5.4.1). Computational parameters employed for the bulk Si and LiF systems within the linear-combination of atomic orbital (LCAO) and planewave projector augmented wave (PAW) simulations carried out in this work are summarized in table~\ref{tab1}.  
\begin{figure}[htbp]
	\centering
	\includegraphics[scale=0.44]{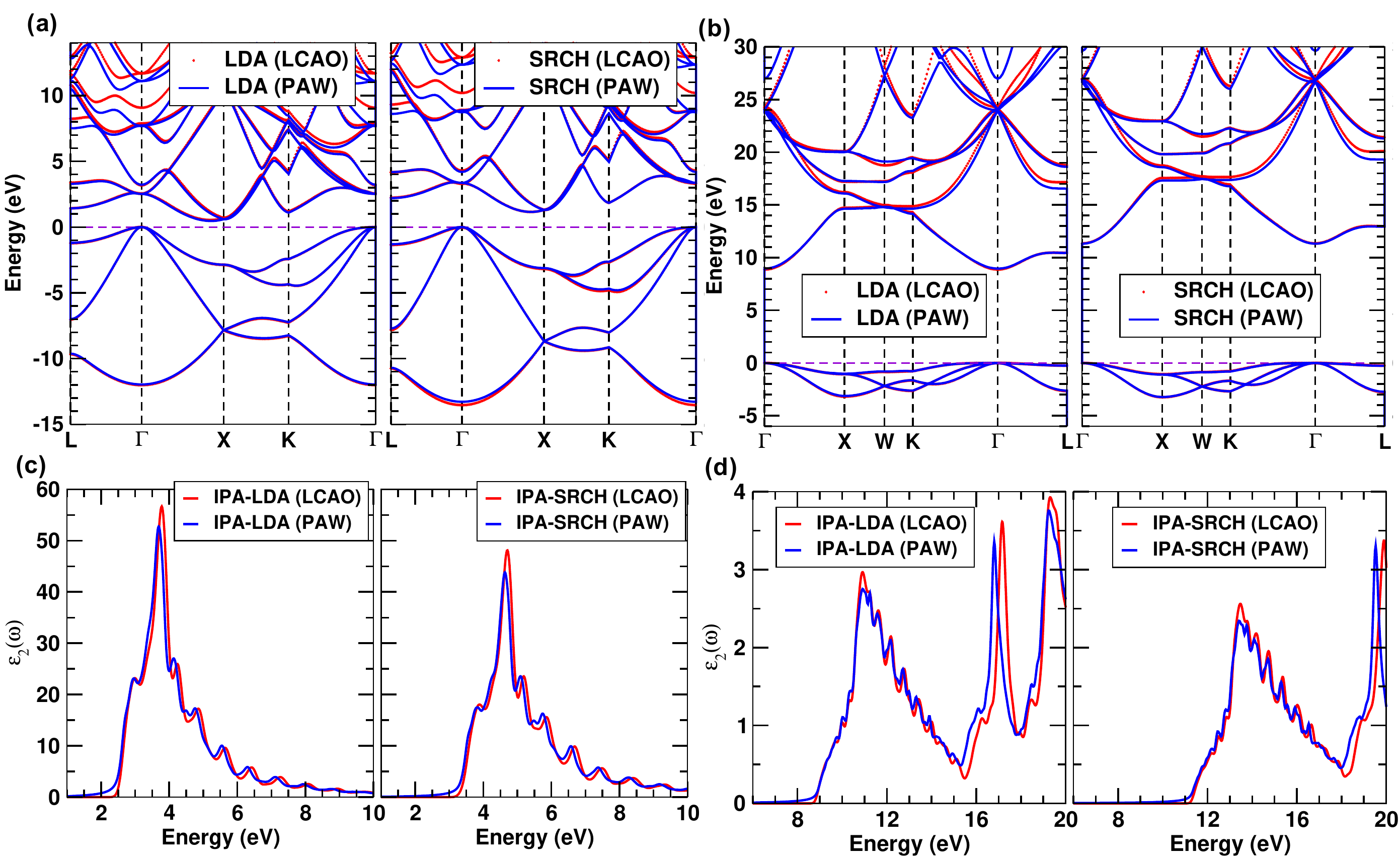}
	\caption{(a,b)~LCAO and PAW band dispersions for bulk Si (a) and LiF (b) obtained at the LDA and short-range-corrected hybrid (SRCH) DFT levels of theory are compared. Parameters characterizing the SRCH functional are listed in table~\ref{prsh}. (c,d) Imaginary parts of the frequency dependent dielectric function for Si (c) and LiF (d) obtained at the LDA and SRCH levels from LCAO and PAW simulations are shown.}
	\label{silif}
\end{figure}
\subsection{Groundstate bandstructures}
In figure~\ref{silif}, band dispersions for bulk Si (Fig.~\ref{silif}(a)) and LiF (Fig.~\ref{silif}(b)) obtained from LCAO and PAW simulations are compared at the LDA and short-range-corrected hybrid (SRCH) DFT levels of theory.  The SRCH functional used here employs the same fraction of short-range HFX and range-separation parameter as the popular HSE06~\cite{Krukau2006} functional. However the semilocal DFT XC components in the SRCH are based on the LDA where as the corresponding quantities within HSE06 are based on the generalized gradient approximation (GGA)~\cite{Krukau2006}. The HFX parameters characterizing the SRCH are reported in table~\ref{prsh}. The equivalent settings within VASP are AEXX=0.25, HFSCREEN=0.2. Several observations can be made with regards to figures~\ref{silif}(a,b): LCAO and PAW band dispersions at the LDA level exhibit close agreement within the valence band (VB) and upto a few eV into the conduction band (CB). At higher energies into the CB, noticeable differences appear as the LCAO basis sets have less variational freedom compared to planewaves. Importantly, band-structures obtained using the SRCH functional also show a similar level of agreement as those at the LDA level especially near the Fermi energy. Numerical values of the band gaps in Si and LiF obtained at the LDA and SRCH DFT levels are also reported in table~\ref{bgap} and show reasonably good agreement. At the SRCH DFT level, the bottom of the VB in bulk-Si comes out slightly more dispersive in the LCAO result, being 0.24 eV lower in energy than the PAW result of -13.28 eV at $\Gamma$. Elsewhere in the VB and within the first 5 eV into the CB, in both Si and LiF, LCAO and PAW band energy differences are small enough to be barely noticeable. This suggests firstly that the real-space GKS DFT implementation in the LCAO code is consistent with a standard PAW implementation for SRCH functionals and furthermore that the LCAO basis sets which are constructed at the LDA level have sufficient flexibility to capture electronic structure changes induced by adopting SRCH functionals. A related perspective is provided by looking at the imaginary part of the frequency-dependent linear dielectric function $\epsilon_2(\omega)$ calculated within the independent particle approximation (IPA). As shown in figure~\ref{silif}(c,d) for the LCAO basis sets employed here, a consistent level of agreement with PAW results is obtained for calculated IPA-$\epsilon_2(\omega)$ line-shapes at both the LDA and SRCH DFT levels.
\begin{figure}[htbp]
	\centering
	\includegraphics[scale=0.46]{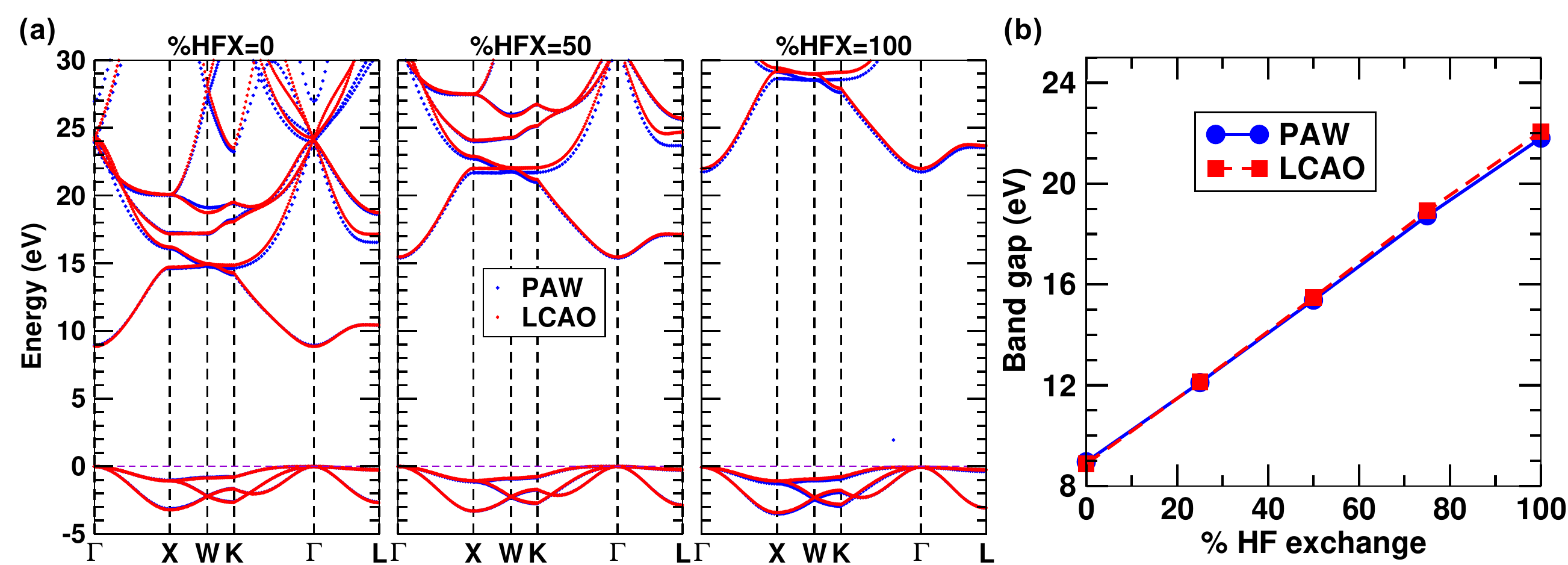}
	\caption{(a) Band dispersions of LiF calculated using global hybrid functionals featuring 0\%,50\% and 100\% Hartree-Fock exchange (HFX). Results from LCAO (red) and PAW (blue) simulations are plotted together.  (b) Variation of the calculated band gap in LiF as a function of \%HFX also comparing LCAO and PAW results.}
	\label{ghlif}
\end{figure}
\begin{table}[htbp]
	\begin{center}
		\begin{tabular}{ |c|c|c|c| }
			\hline
			&Si &LiF&2D h-BN\\
			\hline
			\multirow{3}{*}{SRCH} & $\alpha$=0.25&  $\alpha$=0.25&  \\
			& $\beta$=$-$0.25&  $\beta$=$-$0.25 & - \\
			& $\omega$=0.2~$\mathrm{\AA}^{-1}$&  $\omega$=0.2~$\mathrm{\AA}^{-1}$ &  \\
			\hline
			\multirow{3}{*}{LRCH} & $\alpha$=0.2&  $\alpha$=0.2&  $\alpha$=0\\
			& $\beta$=$-$0.1167&  $\beta$=0.326 &  $\beta$=1\\
			& $\omega$=0.208~$\mathrm{\AA}^{-1}$&  $\omega$=0.695~$\mathrm{\AA}^{-1}$ &  $\omega$=0.238~$\mathrm{\AA}^{-1}$\\
			\hline
			\multirow{3}{*}{GH} & &  &  $\alpha$=0.5\\
			&- & - &  $\beta$=0\\
			& & &  $\omega$=0~$\mathrm{\AA}^{-1}$\\
			\hline
		\end{tabular}
	\end{center}
	\caption{Parameters characterizing the range-separated hybrid DFT functionals employed in this study. The fraction of HFX in the short-range is indicated by $\alpha$ while the fraction of long-range HFX  in each case is given by ($\alpha+\beta$). The short-range-corrected hybrid (SRCH) functionals therefore only include HFX in the short-range. The long-range-corrected hybrid (LRCH) functionals (LRC) include a non-vanishing fraction of long-range HFX and the global hybrid (GH) functional includes the same fraction of HFX at all length-scales.}\label{prsh}
\end{table}
\begin{table}[htbp]
	\begin{center}
		 \begin{adjustbox}{width=1\textwidth}
		\begin{tabular}{ |c|c|c|c|c|c|c|c| }
			\hline
			&\multicolumn{2}{|c|}{LDA} &\multicolumn{2}{|c|}{SRCH}&LRCH&GH&Reference\\
			\hline
			&LCAO&PAW&LCAO&PAW&LCAO&LCAO& \\
			\hline
			Si&0.538&0.463&1.151&1.154&1.191&-&1.17~\cite{Lautenschlager1987} [Expt.]\\
			LiF&8.870&8.961&11.345&11.325&14.229&-&14.2$\pm$0.02~\cite{Piacentini1976} [Expt.]\\
			{2D h-BN}&4.588&-&-&-&7.819&9.046&7.77~\cite{Ferreira2018}[GW]\\
			\hline
		\end{tabular}
	\end{adjustbox}
	\end{center}
	\caption{Band gaps in eV obtained for different choices of GKS functionals are shown for the solid-state materials considered in this study. For Si and LiF band gaps calculated at the LDA and SRCH DFT levels using LCAO basis sets are compared against planewave PAW results. For 2D h-BN the direct band gap at the K point in the Brillouin zone is shown as obtained from the LCAO method using LDA, LRCH and GH functionals in comparison with a literature GW value~\cite{Ferreira2018} in the last column. Parameters characterizing the different functionals are shown in table~\ref{prsh}. Reference values for the band gaps taken from the literature are shown in the last column.}\label{bgap}
\end{table}
Next the inclusion of non-zero long-range HFX exchange in the LCAO implementation is investigated in comparison with the PAW approach by adopting a global-hybrid (GH) functional form for the XC energy where by the same non-zero fraction of HFX operates at all length-scales. This is achieved within the LCAO scheme by setting $\beta=0, \omega=0$ in equations~\ref{ERSH},\ref{RSHC} and choosing $\alpha$ to be between 0 and 1.  The equivalent choice in VASP would be to choose AEXX=$\alpha$ and HFSCREEN=0~\cite{Marsman2008,Hafner2008}. It is also worth noting in this instance that as shown in table~\ref{tab1}, the size of the real-space auxiliary supercell and the reciprocal space $\mathbf{q}$-point grid used for calculating HFX are consistent between the LCAO and PAW approaches respectively. In figure~\ref{ghlif} band structures for LiF obtained from LCAO and PAW simulations employing 0\%, 50\% and 100\% global HFX are compared. Once again, satisfactory agreement is demonstrated in that any discrepancies between LCAO and PAW results are a small compared to the magnitude of the changes induced in the band structures by varying the fraction HFX. Band gaps of LiF obtained from the two approaches are plotted as function of \%HFX in figure~\ref{ghlif}(b). The largest difference between LCAO and PAW band gaps is 0.252 eV observed at 100\% HFX, when, the PAW predicted band gap is 21.807 eV. Thus the LCAO band gap is within 1.2\% of the PAW value across the \%HFX parameter range and given that significantly lower fractions of HFX are needed in practically useful simulations, the discrepancies in practice are expected to be smaller. To summarize the above discussion, for both SRCH and GH DFT functionals the present LCAO implementation is able to reproduce band dispersions that are in agreement with a standard PAW framework to within a few percent.  The validity of the LCAO scheme for general long-range-corrected hybrid (LRCH) DFT functionals featuring a non-zero fraction of long-range HFX follows from the fact that LRCH functionals of the form implied by equation~\ref{ERSH}, can be constructed as linear combinations of SRCH and GH functionals. 

\subsection{Time-domain simulations}
In this subsection generalized Kohn-Sham (GKS) VG-RT-TDDFT~\cite{Baer2018,Sato2015} results obtained using the LCAO implementation are presented. It is noted in this context that a precursor to this implementation has been benchmarked against established real-space grid based VG-RT-TDDFT showing good agreement at the ALDA level of theory both for linear-response and strong-field induced dynamics~\cite{Pemmaraju2018}. Therefore in the following, the emphasis is primarily on the results generated by extending the framework to the GKS level of theory.
\begin{figure}[htbp]
	\centering
	\includegraphics[scale=0.56]{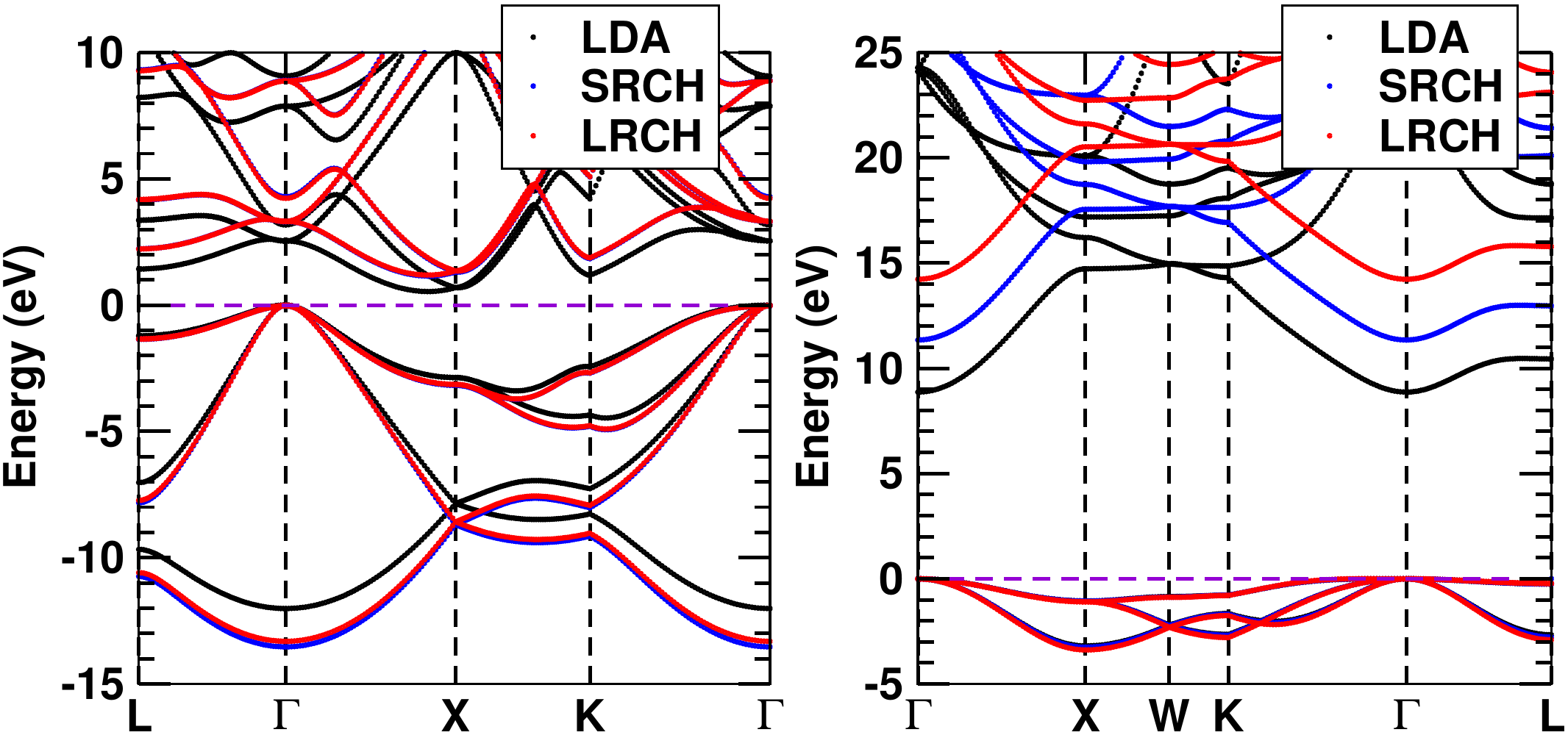}
	\caption{Band structures of Si (left) and LiF (right) obtained using the long-range-corrected hybrid (LRCH) functional employed in subsequent VG-RT-TDDFT simulations are shown in comparison with those at the LDA and SRCH DFT levels. Note that in Si, the SRCH and LRCH functionals predict very similar band-structures where as in LiF the LRCH functional leads to a significant band gap increase. See table~\ref{prsh} for parameters characterizing the different functionals in each system.}
	\label{bslrch}
\end{figure}
\subsubsection{Long-range-corrected hybrid functionals}
In the subsequent discussion, LRCH functionals are investigated within the VG-RT-TDDFT formulation described in the previous section. The main focus is their performance with regards to describing exciton binding in solids. The $\alpha$, $\beta$ and $\omega$ parameters characterizing  LRCH functionals in this study are chosen according to the general arguments put forth in previous works on excitons in solids based on frequency-domain linear-response TDDFT~\cite{Yang2015,Refaely-Abramson2015a}.  Firstly, since the long-range decay of the Coulomb interaction in a dielectric medium is of the form $\frac{1}{\epsilon_{\infty} r}$, the fraction of long-range asymptotic HFX determined by the sum $(\alpha + \beta)$, is set to match $\frac{1}{\epsilon_{\infty}}$. Here $\epsilon_{\infty}$, which represents the macroscopic dielectric constant characterizing the ion-clamped electronic screening response in the static limit, can be chosen either from experiment or first-principles results.  Once $(\alpha + \beta)$ is constrained, as far as valence electronic structure is concerned, $\alpha$ which determines the short-range HFX fraction can be chosen on a heuristic basis to be around 0.2-0.25. Finally, having chosen $\alpha$ and $\beta$ the range-separation parameter $\omega$ is determined by requiring the single-particle GKS band-gap predicted by the LRCH functional to match the fundamental gap from first-principles GW calculations or experiment. The above prescription for LRCH functionals therefore ensures that both the fundamental single-particle band gap and qualitative long-range screening behavior are correctly accounted for and on this basis, TDDFT is able to describe electron-hole excitations satisfactorily in solids~\cite{Refaely-Abramson2015a}. The $\alpha$, $\beta$, $\omega$ parameters chosen for the LRCH functionals in this work are shown in table~\ref{prsh} and the band-structures of Si and LiF obtained using these functionals within the LCAO framework are shown in figure~\ref{bslrch}. Corresponding values for fundamental band gaps are reported in table~\ref{bgap} . From figure~\ref{bslrch} it is apparent that in semiconductors like bulk Si, where functionals such as HSE06~\cite{Krukau2006,Paier2006,Paier2008} and the related SRCH functional employed here already predict accurate single-particle band-dispersions, the LRCH functional does not lead to further significant changes in band structure. This is achieved by compensating for the additional non-zero long-range HFX in the LRCH functional through a smaller fraction of short-range HFX relative to the SRCH case (see table ~\ref{prsh}). On the other hand, in wide gap insulators like LiF, SRCH functionals are known to underestimate the band gap~\cite{Paier2006,Paier2008} where as by construction, the LRCH functional induces further corrections to reproduce the fundamental gap. 
\begin{figure}[htbp]
	\centering
	\includegraphics[scale=0.56]{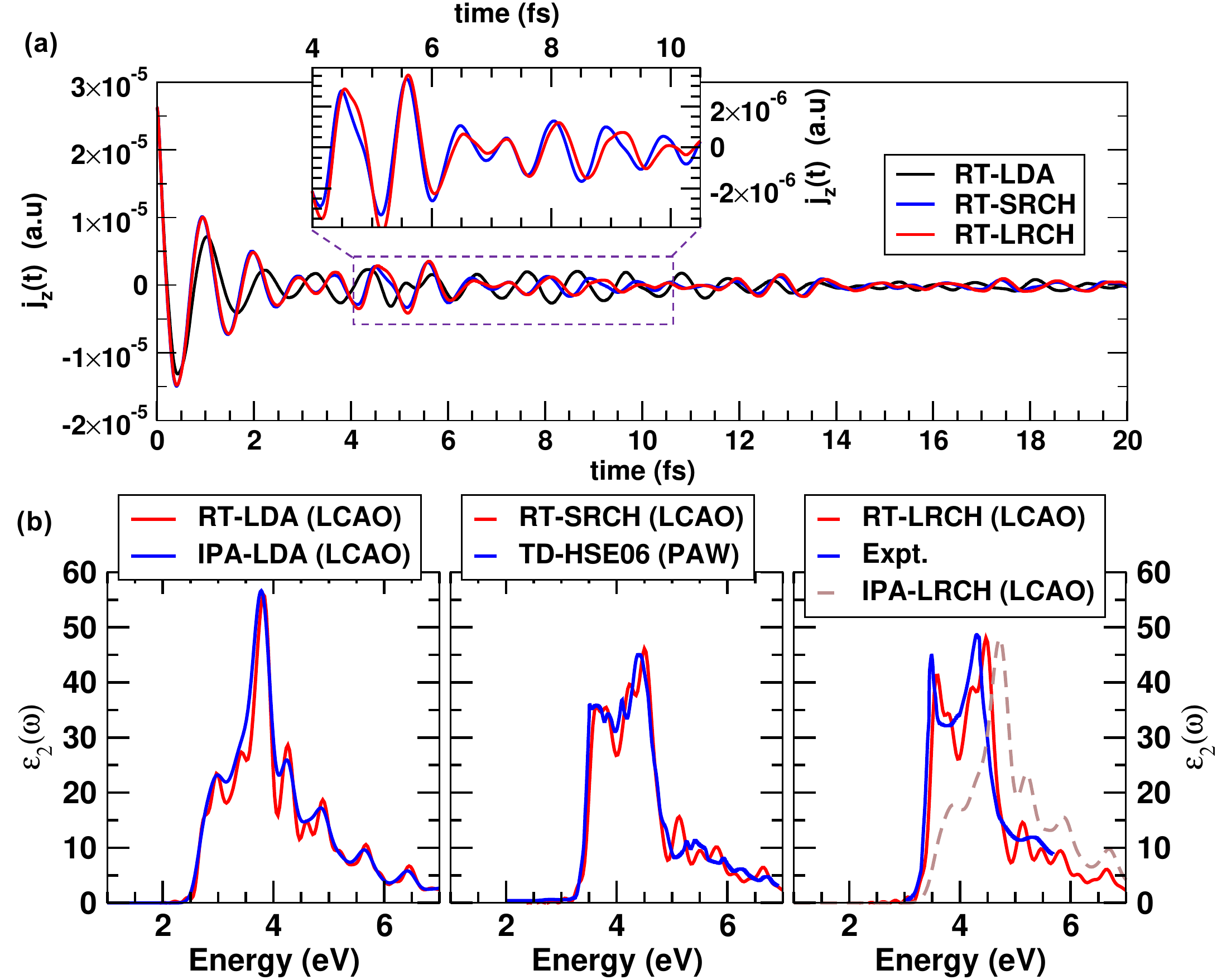}
	\caption{(a) Time dependence of the current induced in bulk Si by a 0.001 a.u delta-function electric field pulse applied at time $t$=0. Results obtained from time-evolution employing different functionals are shown. (b) Imaginary part of the frequency dependent linear dielectric function $\epsilon_2(\omega)$ corresponding to the time-dependent currents shown in (a). (lower-left) $\epsilon_2(\omega)$ from LCAO RT-LDA and the LDA independent particle approximation (IPA)  are compared. (lower-middle) The LCAO RT-SRCH result from this work is compared with the TD-HSE06 result from~reference~\citenum{Paier2008}. (lower-right)  The LCAO RT-LRCH result from this work is compared with experimental data from reference~\cite{Lautenschlager1987}. The LRCH IPA result is also shown.}
	\label{rt-si}
\end{figure}
\subsubsection{Linear optical response in bulk Si}
 In figure~\ref{rt-si}, the time-dependent current in bulk-Si in response to a 0.001 a.u delta function electric-field pulse applied at time $t$=0 is shown as obtained from VG-RT-TDDFT simulations employing LDA, SRCH and LRCH functionals. At very early times the induced current is similar in all cases but differences start to emerge roughly around 1~fs. Primarily, RT-SRCH and RT-LRCH exhibit a shorter time-period for the current oscillations relative to RT-LDA derived from the higher energy onset for optical excitations with the GKS functionals. Furthermore, as shown in the inset within figure~\ref{rt-si}(a), the currents from RT-SRCH and RT-LRCH also exhibit some differences as time progresses. The imaginary part of the frequency dependent dielectric function $\epsilon_2(\omega)$ obtained from these simulations is shown in figure~\ref{rt-si}(b). As expected, the RT-LDA response is very similar to that obtained at the independent-particle approximation (IPA) level of theory. RT-SRCH meanwhile accounts for some fraction of excitonic effects in bulk-Si producing intensity enhancement near the onset of optical absorption around $\sim$3.5 eV. Importantly,  in figure~\ref{rt-si}(b) the RT-SRCH response obtained from the present LCAO scheme is compared against previously published linear-response TD-HSE06~\cite{Paier2008} results in the literature and shows good agreement for the overall line-shape. In the rightmost panel of  figure~\ref{rt-si}(b), the optical response from RT-LRCH is compared against experimental data~\cite{Lautenschlager1987} also showing satisfactory agreement. In particular, the peaks at $\sim$3.5 eV and $\sim$4.3 eV observed in the experimental $\epsilon_2(\omega)$ are reproduced to within 0.2 eV largely consistent with previous LR-TDDFT results~\cite{Refaely-Abramson2015a}. For completeness, the independent-particle  $\epsilon_2(\omega)$ obtained with the LRCH functional is also shown, clearly demonstrating the significant role of excitonic binding in going from IPA to RT-LRCH in bulk Si. 
\begin{figure}[htbp]
	\centering
	\includegraphics[scale=0.6]{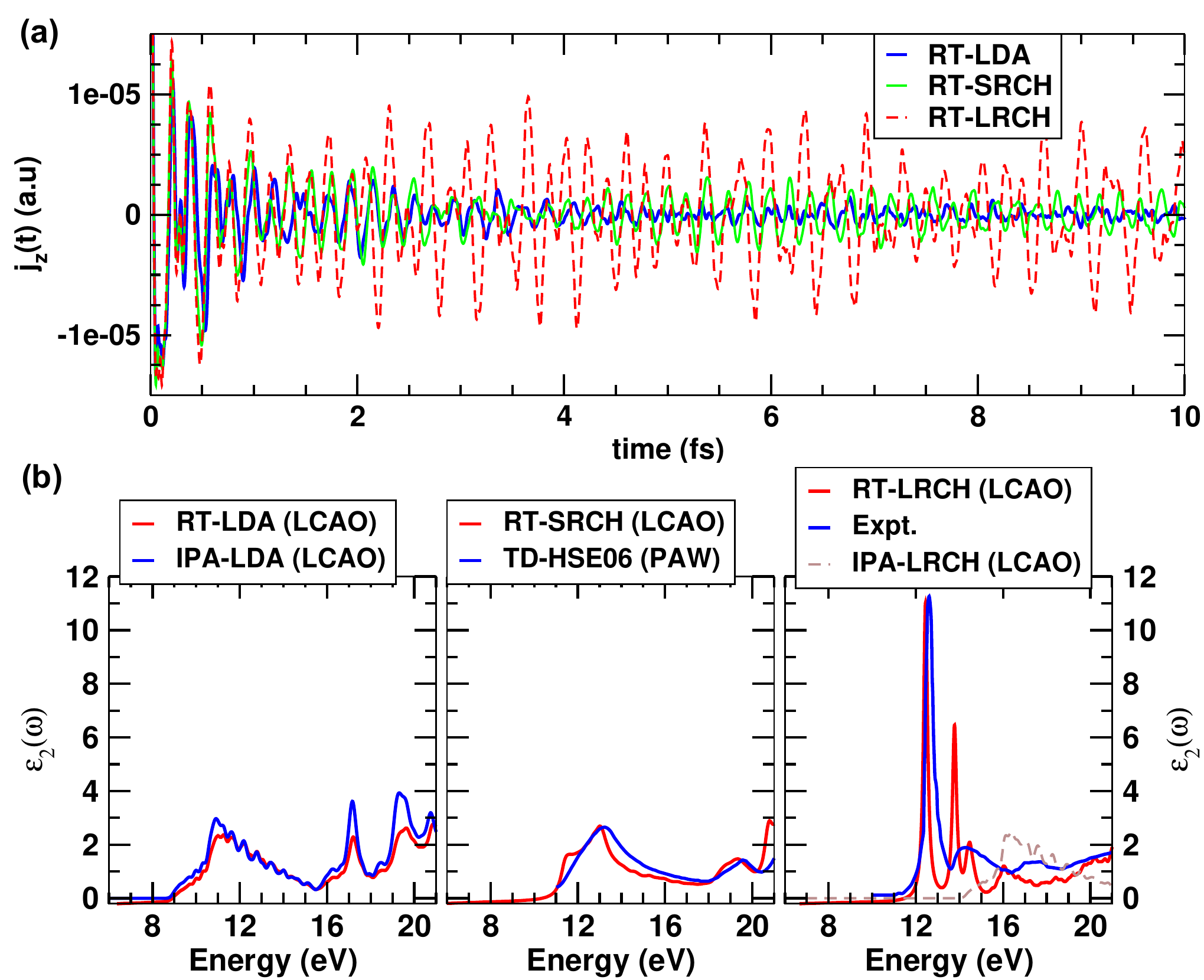}
	\caption{(a) Time dependence of the current induced in bulk LiF by a 0.001 a.u delta-function electric field pulse applied at time $t$=0. Results obtained from time-evolution employing different functionals are shown. (b) Imaginary part of the frequency dependent linear dielectric function $\epsilon_2(\omega)$ corresponding to the time-dependent currents shown in (a). (lower-left) $\epsilon_2(\omega)$ from LCAO RT-LDA and the LDA independent particle approximation (IPA)  are compared. (lower-middle) The LCAO RT-SRCH result from this work is compared with the TD-HSE06 result from~reference~\citenum{Paier2008}. (lower-right)  The LCAO RT-LRCH result from this work is compared with experimental data from reference~\citenum{Piacentini1976}. The LRCH IPA result is also shown }
	\label{rt-lif}
\end{figure}
\subsubsection{Linear optical response in bulk LiF}
In figure~\ref{rt-lif}, results for the more extreme case of the wide gap insulating ionic solid LiF are shown. Figure~\ref{rt-lif}(a) plots the time-dependent current in GKS-VG-RT-TDDFT induced by a small 0.001 a.u delta-function electric-field pulse applied at t=0 to a bulk unitcell of LiF.  Results for RT-LDA, RT-SRCH and RT-LRCH time-propagation are compared. As in the case of bulk Si, at very short times under $\sim$ 1 fs, the induced current looks similar with all three functionals. Subsequently, the RT-SRCH and RT-LRCH currents deviate from the RT-LDA one. Nevertheless, the RT-SRCH and RT-LDA currents look qualitatively similar in that the current oscillations exhibit a significant amplitude reduction at later times due to dephasing. In contrast, the RT-LRCH current differs markedly starting from around $\sim$1.5 fs and exhibits the clear onset of a strong quasi-monochromatic oscillation that undergoes very little amplitude reduction over the time-frame shown. The imaginary part of the frequency dependent dielectric function $\epsilon_2(\omega)$ derived from the above time-propagation is plotted in figure~\ref{rt-lif}(b). Once again, the RT-LDA result closely resembles that of the IPA showing no exciton binding. The RT-SRCH $\epsilon_2(\omega)$  shown in the middle panel of ~\ref{rt-lif}(b) predicts a higher onset of absorption but is otherwise qualitatively similar to RT-LDA.  As is already well known from previous studies~\cite{Paier2008,Yang2015,Refaely-Abramson2015a}, SRCH functionals do not have sufficient long-range nonlocal exchange to reproduce the strongly-bound Frenkel exciton in LiF. The LCAO result obtained for the RT-SRCH functional is in reasonable agreement with the planewave TD-HSE06 reported previously~\cite{Paier2008}. Note that the TD-HSE06 results from reference~\citenum{Paier2008} are averaged over several different $\mathbf{q}$-point grids and therefore exhibit a smoother line shape than the LCAO results which correspond to a single 8$\times$8$\times$8 real-space auxiliary cell. Nevertheless this auxiliary supercell size is consistent with the fine $\mathbf{q}$-point grid used for LiF in recent optimally-tuned RSH LR-TDDFT simulations~\cite{Refaely-Abramson2015a}. In contrast to RT-LDA and RT-SRCH, the RT-LRCH result shown on the right of~\ref{rt-lif}(b) reproduces the qualitatively different optical response measured experimentally~\cite{Piacentini1976}. The pronounced quasi-monochromatic oscillations characterizing the time-dependent current in the RT-LRCH simulation translate in the frequency domain to an intense excitonic peak at 12.45 eV which is $\sim$0.15 eV from the experimental peak at $\sim$12.6 eV~\cite{Piacentini1976}. Besides the main excitonic peak, two smaller peaks at 13.78 eV and 14.47 eV are also observed in the RT-LRCH $\epsilon_2(\omega)$. Although features at these energies are significantly broadened out in experiment, two peaks at very similar energies are clearly observed also in previous planewave LR-TDDFT results~\cite{Refaely-Abramson2015a,Yang2015}. Nevertheless, within the LCAO basis set, the intensity of the second peak near $\sim$13.78 eV seems to be overestimated relative to planewave LR-TDDFT~\cite{Refaely-Abramson2015a,Yang2015}. To summarize the discussion on bulk solids, Si and LiF, the LCAO VG-RT-TDDFT results for both short-range-corrected and long-range-corrected hybrid functionals are generally consistent with equivalent planewave results or experiment and successfully describe the qualitative improvements afforded by the GKS framework in comparison to conventional KS ALDA. Quantitative agreement with respect to excitonic peak positions is within 0.2 eV of reference values. 

\begin{figure}[htbp]
	\centering
	\includegraphics[scale=0.6]{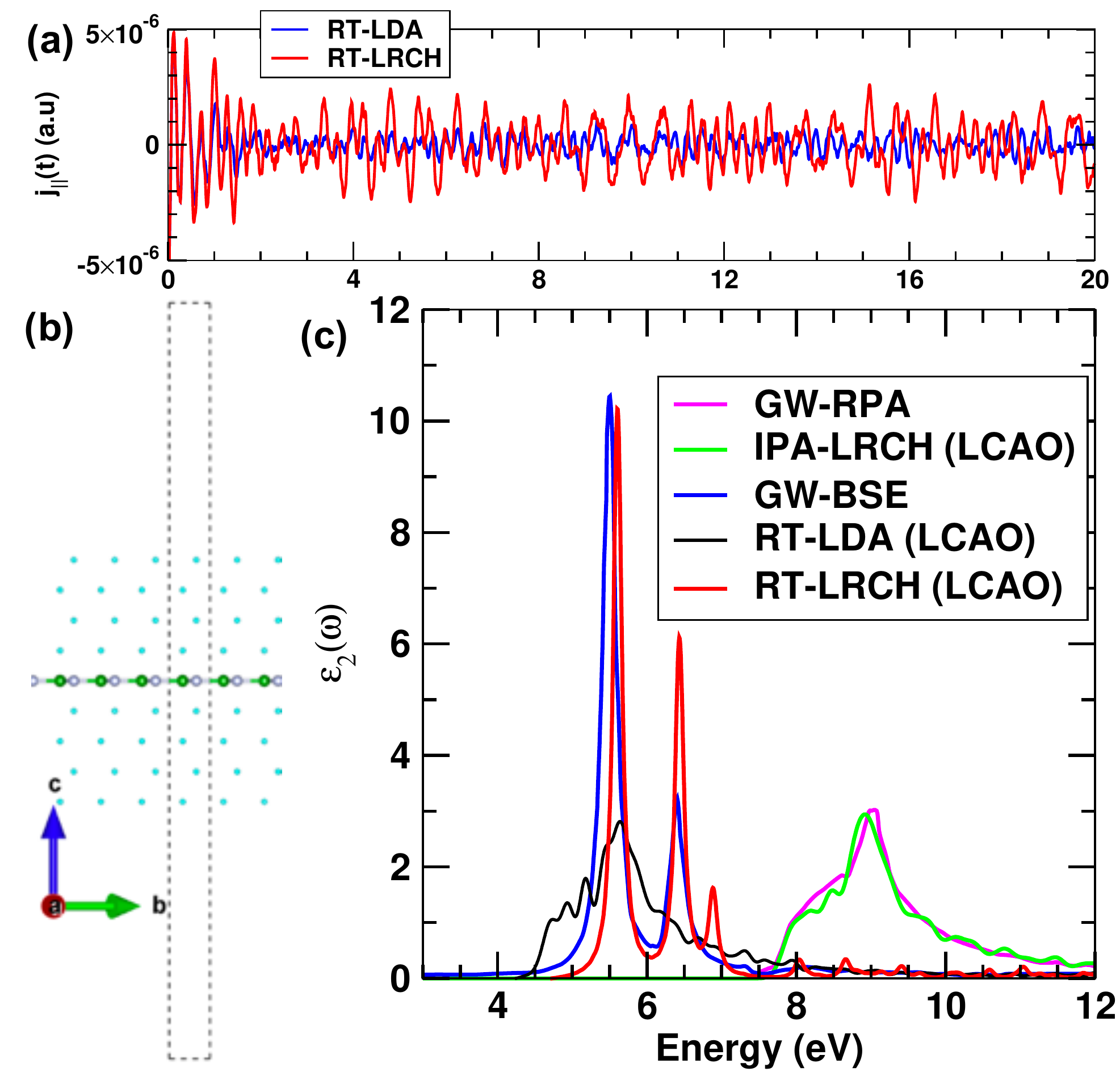}
	\caption{(a) Time dependence of the current induced in monolayer h-BN (2D h-BN) by a 0.001 a.u in-plane-polarized delta-function electric field pulse applied at time $t$=0.  Results obtained at the RT-LDA and RT-LRCH levels are shown. (b) Unitcell (dotted line) employed for VG-RT-TDDFT simulations of 2D h-BN. The light-blue dots represent centers of $ghost$ basis functions included in the vacuum region near the monolayer. (c) Imaginary part of the frequency dependent linear dielectric function $\epsilon_2(\omega)$ corresponding to the time-dependent currents shown in (a) is compared against that from other relevant approximations. The GW-RPA and GW-BSE results are from reference~\citenum{Ferreira2018} }
	\label{rt-hbn}
\end{figure}
\begin{table}[htbp]
	\begin{center}
	 \begin{adjustbox}{width=1\textwidth}
		\begin{tabular}{ |c|c|c|c| }
			\hline
			\multicolumn{4}{|c|}{LCAO computational parameters for 2D h-BN simulations} \\
			\hline
			& no core-states &  B 1$s$ in valence  & N 1$s$ in valence \\
			\hline
			lattice constant (\AA)  &  \multicolumn{3}{|c|}{2.504} \\
			\hline
			LDA pseudopotential & B: [He]2$s^2$,2$p^1$ & B: 1$s^2$,2$s^2$,2$p^1$&B: [He]2$s^2$,2$p^1$\\
			valence configuration & N: [He]2$s^2$,2$p^3$ & N:[He]2$s^2$,2$p^3$&N:1$s^2$,2$s^2$,2$p^3$\\
			\hline
			\multirow{3}{*}{basis set ($nl$-$\zeta$)} &  B: $2s$-2,$2p$-2,$3d$-1  & B: $1s$-2,$2s$-2,$2p$-2,$3d$-1  & B: $2s$-2,$2p$-2,$3d$-1   \\
			&N: $2s$-2,$2p$-2,$3d$-1  & N: $2s$-2,$2p$-2,$3d$-1  & N:$1s$-2,$2s$-2,$2p$-2,$3d$-1   \\
			& 8$\times ghost:2s$-1&  8$\times ghost:2s$-1&  8$\times ghost:2s$-1   \\
			\hline
			Real-space mesh-cutoff& 408 Ry  & \multicolumn{2}{|c|}{940 Ry} \\
			\hline
			HFX auxiliary supercell & \multicolumn{3}{|c|}{$12\times12\times1$}  \\
			\hline			
			DFT SCF $\mathbf{k}$-point grid &\multicolumn{3}{|c|}{$\Gamma-24\times24\times1$}\\
			\hline
			IPA optics $\mathbf{k}$-point grid &$\Gamma-48\times48\times1$&-&-\\
			\hline
			VG-RT-TDDFT $\mathbf{k}$-point grid & \multicolumn{3}{|c|}{$\Gamma-24\times24\times1$}\\
			\hline
			VG-RT-TDDFT time step & 0.08 a.u & \multicolumn{2}{|c|}{0.0075 a.u} \\
			\hline
			\multicolumn{4}{|c|}{Computational parameters for planewave core-hole (CH) DFT approach} \\
			\hline
			& - & B K-edge & N K-edge \\
			\hline
			GGA ultrasoft  & - & B: 1$s^2$,2$s^2$,2$p^1$&B: [He]2$s^2$,2$p^1$\\
			pseudopotential   & - & N:[He]2$s^2$,2$p^3$&N:1$s^2$,2$s^2$,2$p^3$\\
		    configuration  & - & B$_\mathrm{CH}$:1$s^1$,2$s^2$,2$p^4$&N$_\mathrm{CH}$:1$s^1$,2$s^2$,2$p^4$\\
			\hline
			planewave cutoff  &  & \multicolumn{2}{|c|}{25 Ry} \\
			\hline
			supercell dimensions & & \multicolumn{2}{|c|}{15.024$\times$15.024$\times$16 $\AA^3$}  \\
			\hline
			DFT SCF $\mathbf{k}$-point grid & &  \multicolumn{2}{|c|}{$\Gamma$-2$\times$2$\times$1}  \\
			\hline
			Fine $\mathbf{k}$-point grid for XAS& &  \multicolumn{2}{|c|}{$\Gamma-$5$\times$5$\times$5}  \\
			\hline
		\end{tabular}
		\end{adjustbox}
	\end{center}
	\caption{Computational parameters used for valence and core excitation simulations of 2D h-BN. To model the interaction with ionic cores, norm-conserving LDA pseudopotentials generated using the Troullier-Martins~\cite{Troullier1991} scheme are employed within the LCAO code while planewave CH-DFT simulations use ultrasoft~\cite{Vanderbilt1990} GGA pseudopotentials previously benchmarked for h-BN XAS computations~\cite{Huber2015} using the DFT-XCH~\cite{Prendergast2006,Prendergast2009a} approach.  $\Gamma$-centered $\mathbf{k}$-point grids  of the dimensions listed are used to converge different aspects of the simulations. The LCAO basis set is indicated using $nl$-$\zeta$ notation where $n,l$ are principal and azimuthal quantum numbers respectively and $\zeta$ is the number of functions of each $nl$ type. Additional $ghost$ basis functions in the vacuum region surrounding the 2D h-BN layer within the LCAO approach are also indicated.}\label{hbncomp}
\end{table}
\subsubsection{Valence optical response in monolayer hexagonal BN}
The final material system considered in this article is monolayer hexagonal-BN (h-BN) which is a paradigmatic two-dimensional (2D) material with characteristically different screening properties compared to bulk solids~\cite{Thygesen2017a}. Both low energy valence excitations in the UV and high energy core excitations in the soft X-ray range are considered in this context. In 2D materials the underscreening of the Coulomb interaction leads to strong exciton binding on the order of  1 eV even when the fundamental band gap is not very large~\cite{Thygesen2017a}. According to recent first-principles many-body perturbation theory (MBPT) results based on the GW+BSE\cite{Hybertsen1986,Rohlfing2000} approach,  2D h-BN features a $\sim$7.77 eV gap at the K point in the Brillouin zone (BZ) along with a large exciton binding energy of $2.19$ eV~\cite{Ferreira2018}. Because of its simple structure, chemical composition and interesting excitonic response, 2D h-BN represents a good test platform to investigate the performance of GKS-VG-RT-TDDFT with regards to describing the optical properties of 2D materials. In figure~\ref{rt-hbn}(a), the structure of the unitcell used in the LCAO VG-RT-TDDFT simulations is shown. Because 2D systems are surrounded on either side by vacuum, within LCAO schemes, additional basis functions that are not centered on any specific atom but cover the region near the material are often necessary to adequately describe the exponential decay of the electron density into the vacuum region~\cite{Paier2009}. In this work four layers of such \textit{ghost} orbitals of B-$2s$ character are introduced on either side of the h-BN layer. A vacuum region of 40\AA~is used to separate neighboring h-BN periodic images along the direction normal to the layer. Since HFX is constructed in real-space, any nonlocal exchange interaction between periodic images along the normal is trivially excluded within the current scheme. Other computational parameters employed to model 2D h-BN are listed in table~\ref{hbncomp}. 
The LRCH functional used in this instance is constructed according to the following rationale: Firstly, as discussed previously by Huser et al~\cite{Huser2013}, the long-wavelength (q$_{||}$$\rightarrow$0) dielectric constant in 2D materials tends to unity. This limit is adopted here setting $\epsilon_{\infty}$=1 which therefore implies ($\alpha + \beta$)=1 or 100\% HFX in the long-range. Second, short-range HFX fraction is set to zero, i.e., $\alpha$=0. This choice, determined through trial and error to optimize intermediate-range exchange, deviates from the heuristic value of $\alpha$=0.2 employed above in bulk solids. The range-separation parameter $\omega$ is then tuned to approximate the fundamental band gap in 2D h-BN predicted by GW calculations as listed in table~\ref{bgap}.  In figure~\ref{rt-hbn}(d) the time-dependent current induced in 2D h-BN for a small delta-function electric-field at $t$=0, polarized in the plane of the monolayer is shown at the RT-LDA and RT-LRCH levels of theory. RT-LRCH exhibits clear differences from RT-LDA in the form of stronger current oscillations that do not show significant dephasing. In the frequency-domain this manifests as intense excitonic peaks in the corresponding $\epsilon_2(\omega)$ predicted by RT-LRCH that are in reasonably close agreement with recent GW-BSE results (see figure~\ref{rt-hbn}(c))~\cite{Ferreira2018}. In contrast $\epsilon_2(\omega)$ from RT-LDA closely resembles the lineshape obtained at the level of the IPA completely lacking any exciton binding. Note that in figure~\ref{rt-hbn}(c), the IPA $\epsilon_2(\omega)$ from employing the LRCH functional is compared against the GW-RPA result from reference~\citenum{Ferreira2018}, also showing very good agreement.  Therefore, the parameters chosen for the LRCH functional employed in this instance simultaneously approximate MBPT results for the band gap, the IPA response and excitonic effects though VG-RT-TDDFT.  
\begin{figure}[htbp]
	\centering
	\includegraphics[scale=0.5]{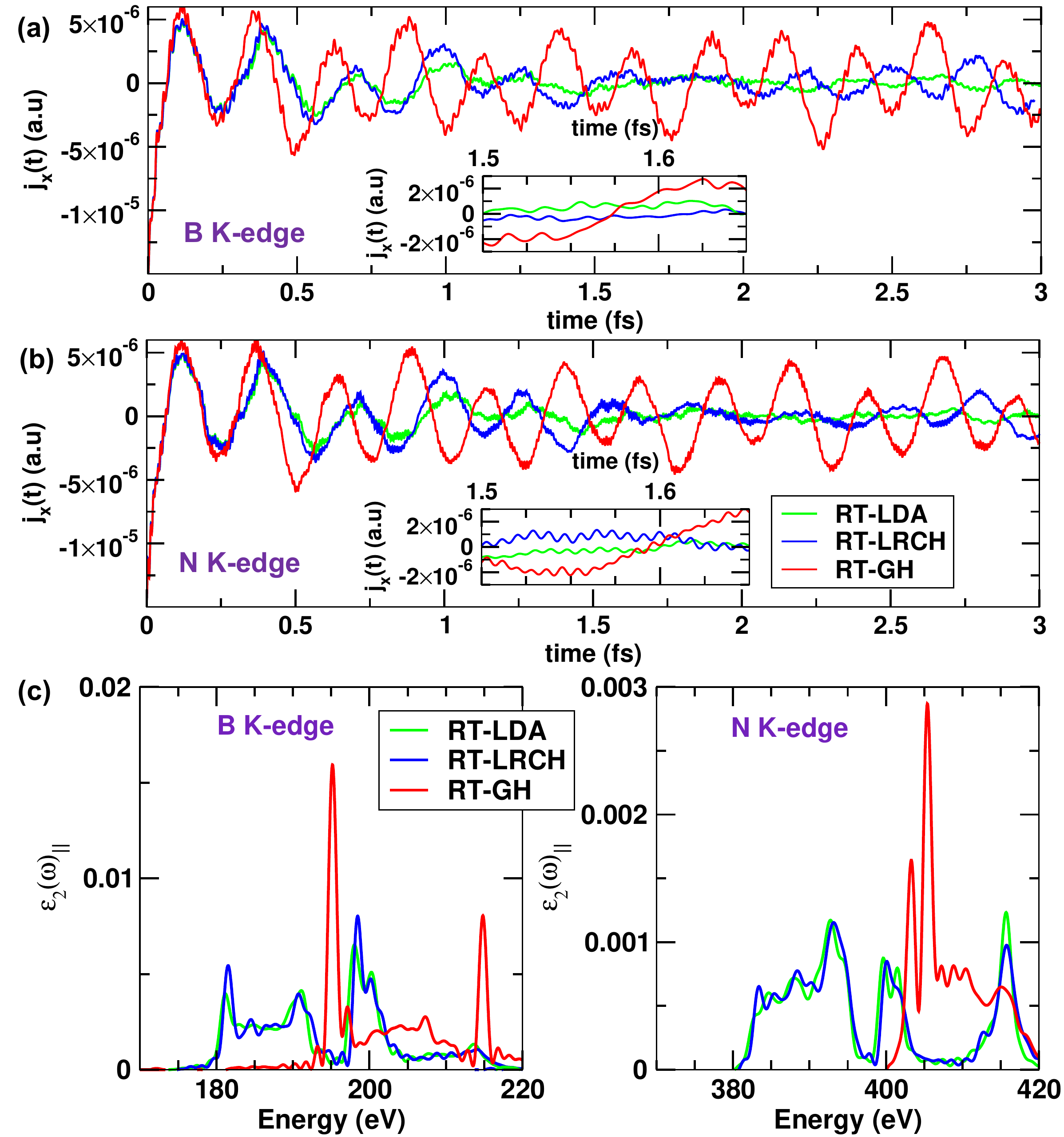}
	\caption{(a,b) Time dependence of the current induced in monolayer h-BN (2D h-BN) by a 0.001 a.u in-plane-polarized delta-function electric field pulse applied at time $t$=0.  Results obtained at the RT-LDA, RT-LRCH and RT-GH levels are shown for the case of B-$1s$ core-states (a) and N-$1s$ core-states (b) included in the valence. (c) Imaginary part of the frequency dependent linear dielectric function $\epsilon_2(\omega)$ corresponding to the time-dependent currents shown in (a,b) is plotted at the B (left) and N (right) K-edge energies.}
	\label{rt-bknk}
\end{figure}
\subsubsection{Core-excitations in monolayer hexagonal-BN}
Having analyzed the performance of GKS VG-RT-TDDFT for valence excitations in both bulk and 2D solids the extension to high energy core-excitations in the soft X-ray energy range is considered next. A comprehensive exploration of core-level spectra in a range of solid-state materials is not undertaken here. Instead, 2D h-BN is employed as a paradigmatic test system to highlight some differences between valence- and core-excitations in the context of VG-RT-TDDFT. Note that the performance of LCAO based VG-RT-TDDFT with respect to describing core-excitations in solids at the ALDA level has been previously investigated in comparison with established FP-LAPW~\cite{Dewhurst2004} methods showing satisfactory agreement~\cite{Pemmaraju2018}. The following discussion therefore proceeds from that baseline. Computational parameters relevant to core-excitation calculations are also listed in table~\ref{hbncomp}. In figures~\ref{rt-bknk}(a,b) the time-dependent current obtained for an in-plane-polarized delta-function electric field perturbation at time $t$=0 is shown for two separate simulations one including B 1$s$ core-states (figure~\ref{rt-bknk}(a))  and the  other including N 1$s$ core-states (figure~\ref{rt-bknk}(b)) in the description.  Three classes of functionals namely RT-LDA, RT-LRCH and RT-GH are considered, the parameters for the latter two being listed in table~\ref{prsh}. Note that the 1$s$ core electrons are explicitly included in the simulation on the same footing as the 2$s$ and 2$p$ electrons mimicking an all-electron description~\cite{Pemmaraju2018}.  In general, excitations from deep-lying core-levels introduce high-frequency modulations of the current superimposed on a slowly varying valence background the latter roughly resembling the current profile obtained if no core-states are included. Insets within figures~\ref{rt-bknk}(a,b) show a magnified view of the current where the core-excitation induced fast oscillations are clearly apparent. Furthermore, the frequency of the fast oscillations is higher for the N K-edge case than the B K-edge case as the former is much higher in energy. 
\begin{table}[htbp]
	\begin{center}
		\begin{tabular}{ |c|c|c|c|c| }
			\hline
			& LDA & LRCH & GH & Expt.~\cite{Trehan1990} \\
			\hline
			B $1s$ & 175.7 & 177.7 & 192.5 & 191 \\
			\hline
			N $1s$ & 376.9 & 378.9 & 400.5 & 398 \\
			\hline
		\end{tabular}
	\end{center}
	\caption{GKS single-particle eigenvalues in eV of the B and N $1s$ core levels obtained by employing the LDA, LRCH and GH functionals (see table~\ref{prsh} for functional parameters) are shown. Experimental XPS core-level binding energies in eV from reference~\cite{Trehan1990} are shown in the last column}\label{xps}
\end{table}
The in-plane component of $\epsilon_2(\omega)$ obtained from the time-dependent current is plotted in figures~\ref{rt-bknk}(c) for the B and N K-edges. It is seen that the RT-LRCH result is very similar to the RT-LDA result showing only small changes in predicted peak intensities and no significant additional exciton binding. Note also that while ALDA is expected to underestimate excitation energies relative to experiment, the absolute energies for the onset of K-edge absorption from RT-LRCH are very similar to RT-LDA.  The absolute GKS eigenvalues relative to the vacuum level, of the N and B 1$s$ core states obtained from groundstate LDA and LRCH DFT are shown in table~\ref{xps}. Clearly, also at the single-particle level, both the LDA and LRCH functionals underestimate the binding energy of 1$s$ core-states relative to X-ray photoemission spectroscopy (XPS) observations~\cite{Trehan1990}. Therefore the LRCH functional that performs well for valence band gaps and exciton binding essentially fails for core-excitations. This is not very surprising given that excitonic effects involving core- and valence-excitations operate on somewhat different length-scales~\cite{Shirley2006,Song2008,Besley2009,Imamura2015}. In core-excitons the highly-localized nature of the core-hole renders short-range screening in its immediate neighborhood very important~\cite{Shirley2006} requiring significant amounts of short-range nonlocal exchange within a GKS description~\cite{Song2008,Besley2009,Imamura2015}. The LRCH functional used in this study features zero HFX in the short-range and therefore is unable to produce significant core-exciton binding. The above phenomenology is already known from molecular core-level spectroscopy using TDDFT where specially designed  short-range-corrected hybrid functionals~\cite{Besley2009} or multiply-range-separated functionals~\cite{Song2008,Imamura2015} have been proposed as a means of improving absolute edge energies in X-ray absorption spectra (XAS). The situation nevertheless seems somewhat more severe in the case of solids. The bound unoccupied single-particle states that are generally responsible for near-edge XAS lineshapes in small molecular systems form discrete well-spaced energy levels regardless of the DFT functional employed and electron-hole attraction effects in small molecules usually manifest as energy shifts but without involving large spectral intensity changes. Therefore, in molecules, LRCH or global hybrid functionals useful for predicting valence electronic structure are also known to usually reproduce satisfactory XAS spectral lineshapes but with the absolute edge energies having to be offset rigidly on the order of $\sim$10 eV to match experiment~\cite{Zhang2015,Petrenko2008,Attar2017}. In solids on the other hand, the conduction band is a continuum of states and strongly bound core- or valence-excitons can lead to qualitatively different spectral lineshapes featuring large near-edge intensity changes and energy shifts with short-range HFX being important to capture either effect.   
\begin{figure}[htbp]
	\centering
	\includegraphics[scale=0.43]{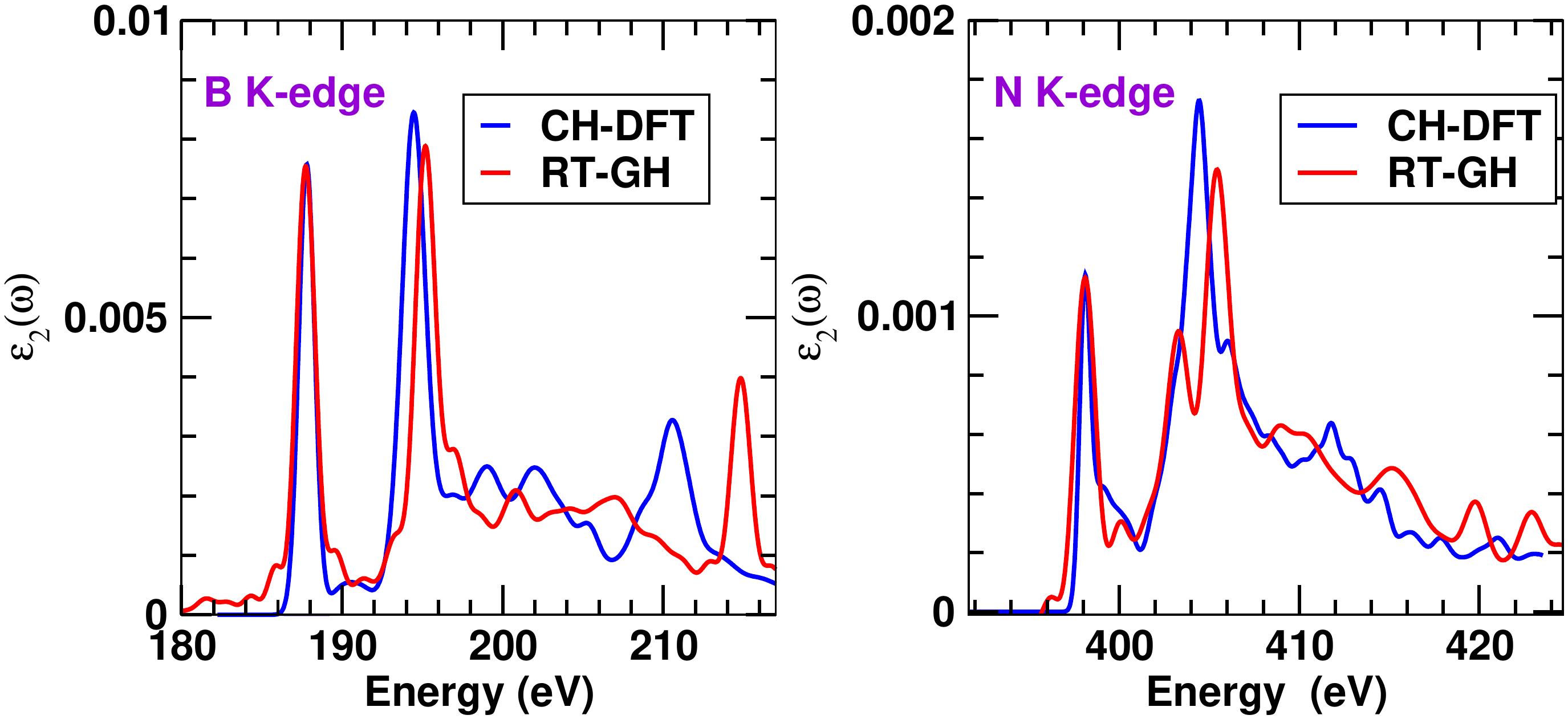}
	\caption{B (left) and N (right) K-edge XAS spectra in monolayer h-BN, obtained from LCAO based VG-RT-TDDFT using a global hybrid functional and a standard planewave CH-DFT method are compared. CH-DFT spectra which do not have an absolute energy scale, are shifted rigidly to match the first peak position from VG-RT-TDDFT in each case.}
	\label{xch}
\end{figure}
Simulating core- and valence-excitation spectra simultaneously, as would be relevant for pump-probe~\cite{Schultze2014} or nonlinear~\cite{Zhang2015} X-ray spectroscopies, requires GKS functionals that balance both long- and short-range HFX and efforts in this direction have been made in the context of molecules through modified range-separation schemes~\cite{Song2008,Imamura2015}. Extension of the present solid-state LCAO framework to such functionals will be considered in the future. For the purposes of this article a global hybrid (GH) functional that features 50\% of HFX is considered instead as a way of striking a compromise between long- and short-range HFX. The valence band gap and core-level single-particle eigenvalues obtained using this GH are reported in tables~\ref{bgap} and~\ref{xps} respectively. The valence band gap is overestimated by $\sim$16\% but the core-level eigenvalues are now in much better agreement with XPS data. The time-dependent current for an in-plane-polarized electric field perturbation and corresponding $\epsilon_2(\omega)$ obtained at the RT-GH level are plotted in figure~\ref{rt-bknk}(a,b) for both the B and N K-edges. It is apparent from figure~\ref{rt-bknk}(c) that RT-GH predicts X-ray absorption onsets that are higher in energy and furthermore, a strong increase in peak intensity derived from excitonic effects near the absorption onset is seen, markedly changing the overall spectral shape relative to RT-LDA and RT-LRCH.  To assess the overall performance of RT-GH TDDFT for XAS, the total $\epsilon_2(\omega)$ including both in-plane and out-of-plane core-level response to small delta-function electric field perturbations is calculated at both the B and N K-edges and plotted in figure~\ref{xch}. The VG-RT-TDDFT results in this instance are compared against those from an established planewave based core-hole (CH) DFT approach for XAS simulations known as the XCH method~\cite{Prendergast2006,Prendergast2009a}. Solid-state CH-DFT approaches~\cite{Taillefumier2002,Prendergast2006,Prendergast2009a} which are widely used in core-level spectroscopy modeling employ modified  pseudopotentials to mimic core-excited atoms within a periodic supercell framework and rely on a DFT self-consistent-field procedure to generate Kohn-Sham eigenstates that are subsequently mapped onto XAS. Computational parameters relevant to CH-DFT are reported in table~\ref{hbncomp}. As is apparent from~figure~\ref{xch}, good agreement between LCAO based RT-GH TDDFT and planewave based CH-DFT is demonstrated for the spectral lineshapes at both the B and N K-edges. Therefore at the RT-GH level of theory, consistent improvements in both core-level single particle energies (table~\ref{xps}) and excitonic effects in XAS (figures~\ref{rt-bknk},\ref{xch}) are realized demonstrating the potential utility of GKS VG-RT-TDDFT also as a method for soft X-ray core-level spectroscopy prediction in solids.

\section{Conclusions and Outlook}\label{conc}
In conclusion, an LCAO framework for enabling velocity-gauge RT-TDDFT simulations in solids at the generalized Kohn-Sham (GKS) level of theory has been presented. Groundstate band structures and linear optical response properties, the latter from velocity-gauge (VG) time-domain simulations, were calculated using the LCAO GKS framework and assessed in comparison with planewave basis set results in a variety of solid-state materials such as Si, LiF and 2D h-BN that feature very different dielectric properties. Both valence single-particle and optical-excitation energies predicted by the LCAO scheme were shown to be in agreement with planewave results and/or experiment with deviations being limited to within a few percent. Furthermore, LCAO core-level spectra obtained at the GKS level of theory were shown to match predictions from established planewave based core-hole DFT techniques. The developments discussed here therefore represent a promising avenue towards routine GKS-RT-TDDFT simulations of excitonic effects in solids with relevance to a wide range of spectroscopic applications.

\section*{Acknowledgements}
The author is grateful to Kazuhiro Yabana and Shunsuke Sato for useful discussions. This work was supported by the U.S. Department of Energy, Office of Basic Energy Sciences, Division of Materials Sciences and Engineering, under Contract No. DE-AC02-76SF00515 through TIMES at SLAC. This research used resources of the National Energy Research Scientific Computing Center (NERSC), a U.S. Department of Energy Office of Science User Facility operated under Contract No. DE-AC02-05CH11231. Core-hole DFT simulations in this work made use of the Shirley-XAS method developed and provided by The Molecular Foundry at Lawrence Berkeley National Laboratory, supported by the Office of Science of the U.S. Department of Energy under Contract No. DE-AC02-05CH11231

\section*{References}

\bibliography{tdrsh.bib}

\end{document}